\documentclass[aps,prb,amsmath,amssymb,superscriptaddress,reprint] {revtex4-1}
\usepackage{graphicx} 
\usepackage{pifont}
\usepackage{hyperref}
\usepackage{float}
\usepackage{bm} 

\usepackage{epstopdf}
\usepackage{braket}
\usepackage{color,comment}
\usepackage{textcomp,gensymb}

\hypersetup{ 
    colorlinks=true, 
    linktoc=all,     
    linkcolor=blue,  
    citecolor =blue,
    urlcolor=blue
} 
\usepackage[T1]{fontenc}
\allowdisplaybreaks
    
\usepackage{etoolbox}

\begin{document}

\title{Non-linear soliton confinement in weakly coupled antiferromagnetic spin chains}

\author{H. Lane}
\affiliation{School of Physics and Astronomy, University of Edinburgh, Edinburgh EH9 3JZ, United Kingdom}
\affiliation{School of Chemistry, University of Edinburgh, Edinburgh EH9 3FJ, United Kingdom}
\affiliation{ISIS Pulsed Neutron and Muon Source, STFC Rutherford Appleton Laboratory, Harwell Campus, Didcot, Oxon, OX11 0QX, United Kingdom}

\author{C. Stock}
\affiliation{School of Physics and Astronomy, University of Edinburgh, Edinburgh EH9 3JZ, United Kingdom}

\author{S.-W. Cheong}
\affiliation{Rutgers Center for Emergent Materials and Department of Physics and Astronomy, Rutgers University, Piscataway, New Jersey 08854, USA}

\author{F. Demmel}
\affiliation{ISIS Pulsed Neutron and Muon Source, STFC Rutherford Appleton Laboratory, Harwell Campus, Didcot, Oxon, OX11 0QX, United Kingdom}

\author{R.~A. Ewings}
\affiliation{ISIS Pulsed Neutron and Muon Source, STFC Rutherford Appleton Laboratory, Harwell Campus, Didcot, Oxon, OX11 0QX, United Kingdom}

\author{F. Kr{\"u}ger}
\affiliation{ISIS Pulsed Neutron and Muon Source, STFC Rutherford Appleton Laboratory, Harwell Campus, Didcot, Oxon, OX11 0QX, United Kingdom}
\affiliation{London Centre for Nanotechnology, University College London,
Gordon St., London, WC1H 0AH, United Kingdom}

\date{\today}

\begin{abstract}
We analyze the low-energy dynamics of quasi one dimensional, large-$S$ quantum antiferromagnets with easy-axis anisotropy, using a semi-classical non-linear 
sigma model. The saddle point approximation leads to a sine-Gordon equation which supports soliton solutions. These correspond to the movement of spatially 
extended domain walls. Long-range magnetic order is a consequence of a weak 
inter-chain coupling. Below the ordering temperature, the coupling to nearby chains leads to an energy cost associated with the separation of two domain walls. 
From the kink-antikink two-soliton solution, we compute the effective confinement potential. At distances large compared to the size 
of the solitons the potential is linear, as expected for point-like domain walls. At small distances the gradual annihilation of the solitons weakens the effective 
attraction and renders the potential quadratic. From numerically solving the effective one dimensional Schr\"oedinger equation with this non-linear confinement potential 
we compute the soliton bound state spectrum. We apply the theory to CaFe$_{2}$O$_{4}$, an anisotropic $S=5/2$ magnet based upon antiferromagnetic zig-zag 
chains. Using inelastic neutron scattering, we are able to resolve seven discrete energy levels for spectra recorded slightly below the N\'eel temperature 
$T_\textrm{N}\approx 200$~K. These modes are well described by our non-linear confinement model in the regime of large spatially extended solitons.
\end{abstract}

\maketitle


\section{Introduction}

Confinement and deconfinement of particles, topological defects or fractionalized excitations are recurring motifs in many areas of physics. A famous example 
is the quark-gluon plasma, which is predicted to form at extremely high temperatures. In this new state of matter the quarks and gluons, 
 which under normal conditions are strongly confined in atomic nuclei, behave as asymptotically free particles\cite{Collins+75}.   Another example of a confinement-deconfinement 
 transition is the Berenskii-Kosterlitz-Thouless transition\cite{Berezinsky71,Kosterlitz+73} in two-dimensional XY magnets that is driven by an unbinding of thermally 
 excited vortex-antivortex pairs. 
 
\begin{figure}[t]
    \centering
    \includegraphics[width=0.95\linewidth]{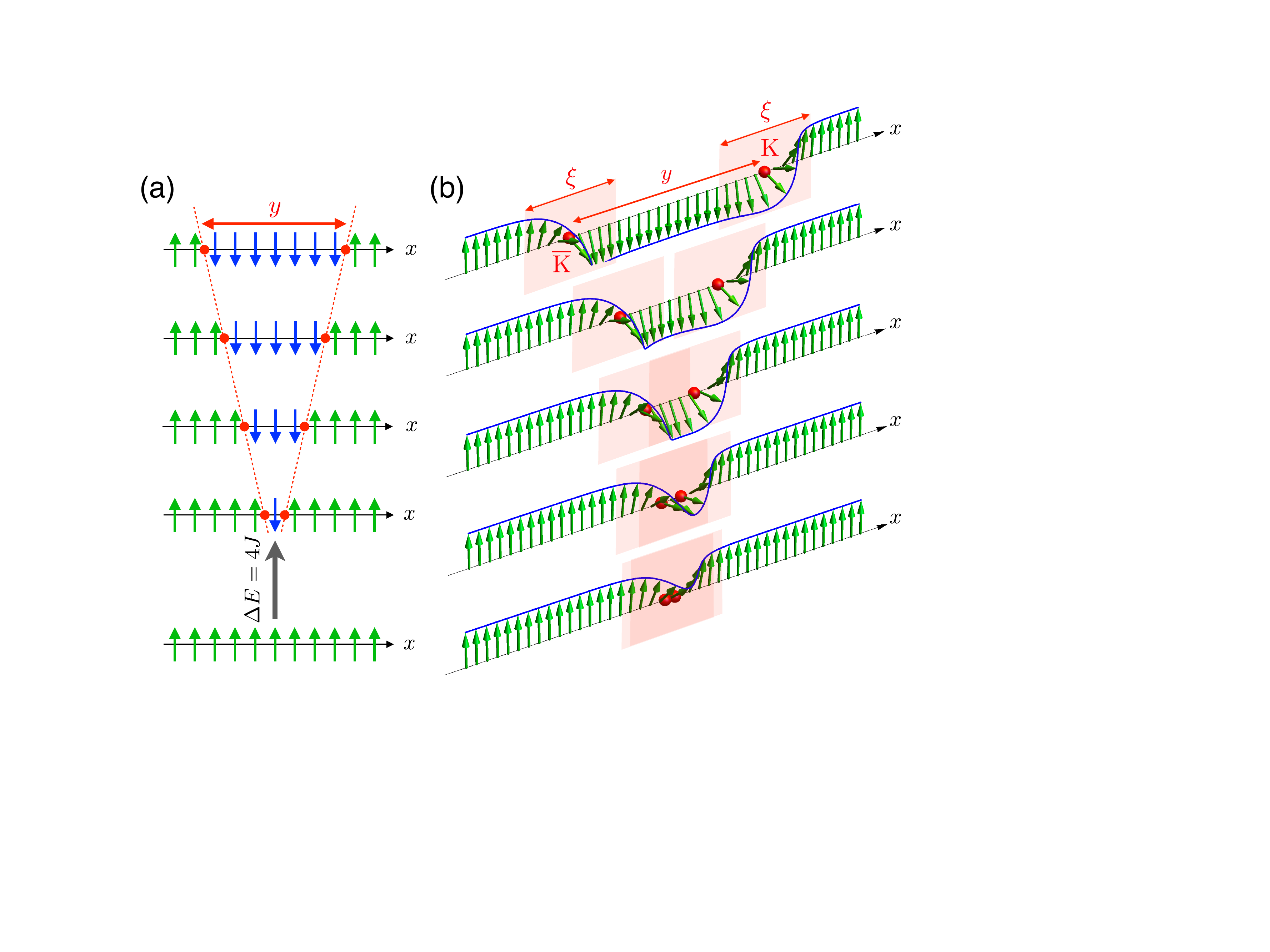}
    \caption{Staggered magnetizations of antiferromagnetic spin chains with Ising anisotropy in the presence of two domain walls (red). (a) For the $S=1/2$ chain, a spin-flip 
    excitation fractionalizes into a pair of spinons. The energy cost due to the coupling to nearby chains scales with the number of spins between the domain walls, 
    giving rise to a linear confinement potential, $V(y)\sim |y|$. (b) For large $S$ spin chains the domain walls are semi-classical chiral solitons of size $\xi$. Shown are different time 
    instances of the collision of a soliton ($\textrm{K}$) and anti-soliton ($\overline{\textrm{K}}$) obtained from the $\textrm{K}\overline{\textrm{K}}$ two-soliton solution of the sine-Gordon 
    equation. The spatial extent of the domain walls causes them to annihilate gradually, rendering the effective confinement potential quadratic at small distances, $V(y)\sim y^2$.}
    \label{fig1}
\end{figure}

Spin-charge separation in one dimension\cite{Tomonaga50,Luttinger,Haldane81} can be viewed as a fractionalization of the electrons into holons and spinons, 
carrying the charge and spin degrees of freedom, respectively. If local repulsions lead to charge localization, the insulating system is well described by the 
antiferromagnetic $S=1/2$ Heisenberg model. In the presence of Ising exchange anisotropy, spinons can be viewed as domain walls in the antiferromagnetic order and are 
created in pairs by a single spin flip (see Fig.~\ref{fig1}(a)). They are therefore fractionalized excitations that carry half of the spin-1 quantum of a magnon 
excitation.\cite{Faddeev+81} If spinons are free to propagate, these pairs are expected to form a triplet excitation continuum. Such continua  are  predicted 
theoretically,\cite{Bougourzi+96,Karbach+97,Caux+08} building on the analytical Bethe Ansatz solution,\cite{Bethe31} and observed experimentally in a number of quasi 
one-dimensional $S=1/2$ antiferromagnets.\cite{Mourigal+13,Bera+17,Gannon+19,Wu+19} 

Staggered $g$-tensors and Dzyaloshinskii-Moriya interactions can lead to an unusual field dependence, such as 
an induced gap,\cite{Oshikawa+97,Affleck+99} $\Delta\sim H^{2/3}$, and field dependent soft modes at incommensurate wave vectors,\cite{Dender+96,Dender+97} as predicted by spinon and 
Bethe Ansatz descriptions.\cite{Pytte+74,Ishimura+77,Muller+81} Through a procedure of bosonization, the dynamics of such systems can be shown to be 
governed by the quantum sine-Gordon model which admits soliton and breather solutions, corresponding to propagating and oscillating domain 
walls, respectively.\cite{Affleck+99,Essler+03} This suggests that spinons can be viewed as quantum solitons \cite{Lake05} and therefore exhibit chirality, 
which was indeed confirmed by polarized neutron scattering.\cite{Braun+05} Soliton and breather modes were identified in 
neutron-scattering \cite{Kenzelmann+04,Umegaki+15} and electron-spin-resonance \cite{Zvyagin+04,Liu+19} experiments.

The effect of a weak interchain interaction is twofold. Firstly, it sets the temperature scale $T_\textrm{N}$ at which two- or three-dimensional long-range order develops.
Secondly, it generates an effective attraction between spinons below $T_\textrm{N}$ since the separation of domain walls will frustrate interchain interactions 
with an associated energy cost that grows linearly with their distance. Such a linear confinement potential gives rise to spinon bound states, leading to a 
quantization of the excitation continuum into discrete energy levels, as observed in BaCo$_2$V$_2$O$_8$,\cite{Grenier+15} 
SrCo$_2$V$_2$O$_8$,\cite{Wang+15,Bera+17} and Yb$_2$Pt$_2$Pb.\cite{Gannon+19} These systems all consist of weakly coupled Ising-Heisenberg 
antiferromagnetic (XXZ) chains of $S=1/2$ moments and the measured spinon bound-state energies are almost perfectly described by the eigenvalues of a 
one-dimensional Schr\"odinger equation with an attractive linear potential.

Linear confinement due to weak interchain coupling is not specific to spinons in $S=1/2$ quantum antiferromagnets but occurs generically for any type 
of kink-like domain-wall excitations. In CoNb$_2$O$_6$, a quasi one-dimensional Ising ferromagnet,  the two-kink continuum breaks up into discrete 
bound-state excitations below the magnetic ordering temperature, with the same characteristic level spacing as in the spinon case.\cite{Coldea+10} 

In this paper we analyze the domain-wall confinement in large-$S$ spin-chain antiferromagnets with easy axis, single-ion anisotropy. Our work is motivated 
by the observation of discrete energy levels in the anisotropic antiferromagnet CaFe$_{2}$O$_{4}$,\cite{Stock+16} a spin-5/2 system consisting of weakly-coupled 
zig-zag chains.  As expected for confinement due to frustrated interchain coupling, the bound states form below the the N\'eel 
temperature $T_\textrm{N}\approx 200$~K.  However, the energy levels do not follow the negative zeroes of the Airy function, as predicted for a linear 
confinement potential.  

In the large-$S$ limit, the low-energy effective field theory of the quantum antiferromagnet is the non-linear $\sigma$ model. Starting from this semi-classical 
description, Haldane demonstrated that the spin dynamics of the one-dimensional quantum antiferromagnet with easy-axis anisotropy is governed
by a sine-Gordon equation which supports soliton solutions.\cite{Haldane83} Hence the domain walls in the antiferromagnetic chain are chiral solitons.  
In these spin textures the staggered magnetization rotates between the two favored orientations in a clockwise or anti-clockwise direction over a typical 
distance $\xi$ (see Fig.~\ref{fig1}(b)). Since the overall chirality in the system is conserved, the domain walls are created in pairs of soliton (kink, $\textrm{K}$) 
and anti-soliton (anti-kink, $\overline{\textrm{K}}$). 

Here we compute the confinement potential $V(y)$ from the $\textrm{K}\overline{\textrm{K}}$ two-soliton solution of the sine-Gordon equation and 
show that the extended nature of semi-classical solitons gives rise to a crossover as a function of the domain-wall separation $|y|$.  At large separations, 
$|y|\gg\xi$, the solitons can be considered as point-like objects, giving rise to a linear confinement potential, $V(y) \sim |y|$. For $|y|<\xi$ the soliton and 
anti-soliton overlap, leading to a gradual annihilation of the defects and preventing the staggered magnetization between domain walls from fully rotating 
to the other easy direction. This reduces the interchain-frustration energy, corresponding to a weakening of the effective confinement potential. We find 
that at small distances, $|y|\ll \xi$, the confinement potential is rendered quadratic, $V(y)\sim y^2$. 

The bound-state spectrum is obtained from the numerical solutions of a one-dimensional Schr\"odinger equation with the computed potential $V(y)$.
Because of the crossover in $V(y)$, the energies of tightly-bound states are almost equidistant, as expected for a harmonic oscillator, while for the weakly-bound 
states at higher energies they approach Airy function behavior as predicted for linear confinement. 

In order to test our theory,  we compare computed spectra to those obtained in inelastic neutron scattering experiments
on high-quality single crystals of CaFe$_{2}$O$_{4}$. Slightly below the N\'eel 
ordering temperature, we are able to resolve seven bound states which are well described by our theory of non-linear confinement of spatially extended 
solitons.

The outline of this paper is as follows. In Sec.~\ref{sec.model} we introduce a generic spin Hamiltonian and resulting low energy, non-linear $\sigma$ model
description of a system of weakly coupled antiferromagnetic chains with single-ion Ising anisotropy. We show that the saddle-point approximation results in a sine-Gordon
equation and briefly review the one and two-soliton solutions. In Sec.~\ref{sec.confinement} we compute the energy of a single spin chain with a pair of domain walls
from the kink-antikink solution, treating the interchain coupling at mean-field level. The bound state energies are obtained from numerical solutions of 
the effective Schr\"odinger equation with the effective non-linear confinement potential. Experimental details and results of our inelastic neutron-scattering experiments on 
CaFe$_{2}$O$_{4}$ are presented in Sec.~\ref{sec.experiments}. We demonstrate  that the measured bound-state energies are well described by our theoretical model. 
Finally, in Sec.~\ref{sec.discussion} we summarize and discuss our results.

\section{Theoretical Model}
\label{sec.model}

Our starting point is a generic spin model of weakly coupled chains with antiferromagnetic Heisenberg couplings $J$ between nearest neighbor along the chains 
and $J_\perp \ll J$ between the chains. Each spin is subject to a single-ion, easy axis anisotropy $\alpha>0$. The Hamiltonian of the system is given by
\begin{eqnarray}
\label{eq.Ham}
\hat{\mathcal{H}} & = & J \sum_{i,m} \hat{\mathbf{S}}_{i,m} \hat{\mathbf{S}}_{i+1,m}-\alpha \sum_{i,m}\left (\hat{S}^{z}_{i,m}\right )^{2}\nonumber\\
& & + J_\perp \sum_{i,\langle m,n\rangle} \hat{\mathbf{S}}_{i,m} \hat{\mathbf{S}}_{i,n},
\end{eqnarray}
where $i$ labels the positions in the chains, $m,n$ the different chains, and $\langle m,n\rangle$ denotes nearest neighbor bonds between adjacent chains. 
In this minimal model, we neglect longer-range exchanges and assume the interchain couplings to be the same in all directions. 
For simplicity, we have neglected exchange anisotropy between different spin components and Dzyaloshinskii-Moriya interactions.
Such terms are not relevant in the case of  Calcium Ferrite ($S=5/2$, $L=0$)  because of the lack of any orbital degrees of freedom. A discussion 
of single-ion anisotropy in systems with quenched orbital moment can be found in Ref.~[\onlinecite{Yosida10}].

\subsection{Non-linear $\sigma$ model}

Let us first focus on an isolated antiferromagnetic chain and drop the chain index for brevity. The effective long-wavelength, non-linear $\sigma$ model is 
obtained using a path integral in imaginary time $\tau\in [0,\beta]$, $\beta = 1/(k_\textrm{B} T)$, and resolving the identities between adjacent time 
slices in terms of over-complete spin-coherent states, $| \mathbf{N}_i(\tau) \rangle$. These states are parametrized by unit vectors $\mathbf{N}_i(\tau)$
and have the property $\langle \mathbf{N}_i(\tau) | \hat{\mathbf{S}}_i | \mathbf{N}_i(\tau) \rangle = S \mathbf{N}_i(\tau)$.

In order to perform a spatial continuum limit, we introduce the staggered N\'eel order-parameter field $\mathbf{n}_i(\tau)$ through the relation 
$\mathbf{N}_i(\tau) = (-1)^i \mathbf{n}_i(\tau) + a \mathbf{L}_i(\tau)$, where $a$ denotes the lattice constant and $\mathbf{L}_i(\tau)$ describes 
the spin fluctuations perpendicular to $\mathbf{n}_i(\tau)$. The latter fluctuations are massive and can therefore be integrated out. After taking
the continuum limit, this procedure leads to the non-linear $\sigma$ model,\cite{Haldane83,Sachdev11,Fradkin13}
\begin{equation}
\label{eq.sigma}
S=\frac{\rho_{S}}{2}  \int_0^\beta \mathrm{d}\tau \int_{-\infty}^\infty \mathrm{d}x\Bigg\{\left(\partial_{x}\mathbf{n}\right)^{2}+\frac{1}{c^{2}}\left(\partial_\tau\mathbf{n}\right)^{2}-\kappa n_{z}^{2} \Bigg\},
\end{equation}
with spin-stiffness $\rho_S$, spin-wave velocity $c$ and easy-axis anisotropy $\kappa$. These parameters are related to the microscopic parameters in the spin 
Hamiltonian (\ref{eq.Ham}),
\begin{equation}
\label{eq.parameters}
 \rho_{S} = JS^{2}a, \quad c = \sqrt{2}JSa, \quad \textrm{and} \;\;  \kappa = \frac{2\alpha}{a^{2}J}.
\end{equation} 

In the absence of anisotropy, $\kappa=0$, the relativistic field theory gives rise to a linear dispersion $\omega = c k$, corresponding to spin-wave excitations of the 
antiferromagnet. This is also reflected by the saddle-point approximation $\delta S/\delta \mathbf{n}(x,t) = 0$ in real time $t=-i\tau$, which gives rise to the classical
wave equation $\partial^2_x \mathbf{n} -\frac{1}{c^2} \partial^2_t \mathbf{n} =0$.

\subsection{Sine-Gordon equation and soliton solutions}

In the presence of anisotropy, it is useful to express the unit vector field $\mathbf{n}(x,\tau)$ in terms of spherical coordinates, 
$\mathbf{n} =(\sin\theta\cos\phi,\sin\theta\sin\phi,\cos\theta)$ since the anisotropy only depends on the polar-angle field $\theta(x,\tau)$,
\begin{eqnarray}
\label{eq.sigma2}
S & = & \frac{\rho_{S}}{2}  \int_0^\beta \mathrm{d}\tau \int_{-\infty}^\infty \mathrm{d}x\Bigg\{ \left(\partial_{x}\theta\right)^{2}+\frac{1}{c^{2}}\left(\partial_\tau\theta\right)^{2}\nonumber\\
& & +\mathrm{sin}^{2}\theta \Big[\left(\partial_{x}\phi\right)^{2}+\frac{1}{c^{2}}\left(\partial_\tau\phi\right)^{2}\Big]-\kappa\, \mathrm{cos}^{2}\theta \Bigg\}.
\end{eqnarray}
The equations of motion are obtained from the saddle-point equations $\delta S/\delta \phi(x,t) = 0$ and $\delta S/\delta \theta(x,t) = 0$. For the azimuthal angle we obtain 
a classical wave equation, $\partial^2_x \phi -\frac{1}{c^2} \partial^2_t \phi=0$. Since we are interested in soliton excitations and not in spin waves we will assume
that $\phi(x,t)=\textrm{const}$. In a system with $z$-axis Ising anisotropy the free energy is independent of the choice of this constant.  This removes all dependence of the 
action (\ref{eq.sigma2}) on $\phi$ and the dynamics for the polar angle is governed by the sine-Gordon equation,
\begin{equation}
\label{eq.sg}
\partial^2_x\theta-\frac{1}{c^2}\partial^2_t \theta    = \frac{1}{2}\kappa\, \textup{sin}(2\theta),
\end{equation}
which is known to admit soliton solutions.\cite{Perring+62,Scharf+92} In terms of dimensionless length and time, 
\begin{equation}
\tilde{x}:= \sqrt{\kappa} x, \quad\textrm{and}\; \tilde{t}:=\sqrt{\kappa} c t,
\end{equation}
the 1-soliton solutions are given by
\begin{equation}
\label{eq.1sol}
\theta_{1,\textrm{K}/\overline{\textrm{K}}}(\tilde{x},\tilde{t}) = 2 \arctan\left[ e^{\pm\gamma(\tilde{x}-\tilde{v}\tilde{t})+\delta_0}\right],
\end{equation}
where $\gamma=1/\sqrt{1-\tilde{v}^2}$ denotes the Lorentz factor and $\tilde{v}=v/c$ the velocity of the relativistic soliton excitation in units of the spin-wave velocity $c$,
which plays the role of the speed of light. The different signs in the exponent correspond to kink ($\textrm{K}$) and antikink ($\overline{\textrm{K}}$), respectively.  
$\delta_0$ is a constant that is determined by the initial conditions. 

New soliton solutions can be generated from known solutions via transformations from one pseudo-spherical surface to another.\cite{Crampin+86} 
By application of such a transformation, known as a B{\"a}cklund transformation, one can generate multiple-soliton solutions from the single soliton.\cite{Rogers+02}
Important for our analysis is the ``kink-antikink" ($\textrm{K}\overline{\textrm{K}}$), 2-soliton solution\cite{Cuenda+11}
\begin{equation}
\label{eq.2sol}
\theta_{2,\textrm{K}\overline{\textrm{K}}}(\tilde{x},\tilde{t}) = 2 \arctan \left[ \frac{\sinh\left( \frac{\tilde{v}\tilde{t}}{\sqrt{1-\tilde{v}^2}}  \right)}{\tilde{v}\cosh\left( \frac{\tilde{x}}{\sqrt{1-\tilde{v}^2}}  \right)}  \right],
\end{equation}
where he have chosen the initial conditions such that $\theta_{2,\textrm{K}\overline{\textrm{K}}}(\tilde{x},\tilde{t}=0)=0$, corresponding to a perfectly ordered chain 
$\mathbf{n}(x)\equiv \hat{\mathbf{e}}_z$ with no defects. The $\textup{K}\overline{\textup{K}}$ solution (\ref{eq.2sol}) therefore describes the creation of a soliton and anti-soliton
at $x=0$ at $t=0$ that propagate outwards in opposite directions for $t>0$. This situation is therefore similar to the creation of two spinons by a single spin flip in the $S=1/2$ 
antiferromagnetic chain. The staggered magnetizations for outwards propagating solitons are shown in Fig.~\ref{fig1} and compared with point-like domain walls.

Another class of 2-soliton solutions that satisfy the sine-Gordon equation (\ref{eq.sg}) are the breathers.\cite{Scharf+92,Cuenda+11} These can be obtained directly from 
the $\overline{\textrm{K}}\textrm{K}$ solution by analytic continuation to imaginary values of the velocity $\tilde{v}$. By doing so, one arrives at the breather solution

\begin{equation}
\label{eq.2solB}
\theta_{2,\textrm{B}}(\tilde{x},\tilde{t}) = 2 \arctan \left[  \frac{\sqrt{1-\tilde{\omega}^{2}}}{\tilde{\omega}}  \frac{\sin\left(\tilde{\omega}\tilde{t}  \right)}{\cosh\left( \tilde{x}\sqrt{1-\tilde{\omega}^2}  \right)}  \right].
\end{equation}
Such semi-classical breathers correspond to two domain walls which oscillate anharmonically within a maximum distance. Crucially, both the breather and $\textrm{K}\overline{\textrm{K}}$ solutions have spatially extended domain walls and so the annihilation of a soliton and an anti-soliton happens gradually (see Fig. \ref{fig1}b).


\section{Soliton Confinement}
\label{sec.confinement}

The theory of linear confinement of spinons in weakly coupled $S=1/2$ antiferromagnetic chains with XXZ-Ising exchange 
anisotropy\cite{Grenier+15,Wang+15,Bera+17,Gannon+19} or of domain walls in quasi one-dimensional Ising ferromagnets\cite{Coldea+10} 
is based on the assumption that domain walls are point-like. In this case, the interchain-frustration energy cost associated with the separation 
of two domain walls is simply proportional to the number of spins $N_y=|y|/a$ between two domain walls with distance $|y|$. This gives rise to a linear confinement
potential $V(y)\simeq J_\perp S^2 n_\perp |y|/a$, where $n_\perp$ denotes the number of neighboring chains and $J_\perp$ is the nearest-neighbor 
interchain coupling. The bound-state spectrum obtained from a one-dimensional Schr\"odinger equation with an attractive linear potential indeed gives a 
convincing description of the experimental data.\cite{Grenier+15,Wang+15,Bera+17,Gannon+19,Coldea+10} 

Here we generalize this approach to describe the non-linear confinement of spatially extended soliton domain walls. Our semi-classical path integral approach allows 
us to treat the finite-width of domain walls and to drop the assumption of Ising alignment. As we will see, in the limit of strong Ising anisotropy, the theory of
linear confinement is recovered.

\subsection{Effective confinement potential}

For a given spin profile along the chain, described by a field $\theta(\tilde{x})$ and constant $\phi(\tilde{x})=\phi_0$, the energy of the chain is given by
\begin{equation}
E_\parallel = \frac{\rho_S \sqrt{\kappa}}{2}\int^{\infty}_{-\infty}\mathrm{d}\tilde{x}\Big\{ \left(\partial_{\tilde{x}}\theta\right)^{2}-\left(\cos^2\theta-1\right)\Big\},
\end{equation}
where we subtracted the energy of a fully polarized chain ($\theta(\tilde{x})\equiv 0$), which diverges in the thermodynamic limit. We treat the the interchain coupling
at mean-field level, introducing the staggered magnetization $M= \left| \langle \hat{S}_{i,m}^z \rangle \right|$. The resulting energy contribution per chain is given by
\begin{equation}
E_\perp = \frac{\rho_{S}}{2\sqrt{\kappa}}g_{\perp}\int^{\infty}_{-\infty}\mathrm{d}\tilde{x}\Big\{1-\cos\theta\Big\},
\end{equation}
where we have again subtracted the contribution for a fully polarized chain and defined the coupling
\begin{equation}
\label{eq.gperp}
g_{\perp}= \frac{2n_{\perp} M J_{\perp}}{a^{2}S J},
\end{equation}
with $n_\perp$ the number of neighboring chains and $J_\perp$ the interchain coupling. Because of the dependence on the magnetic order parameter $M$, the 
coupling $g_\perp$ vanishes above $T_\textrm{N}$.

The effective confinement potential $V(y)$ between a soliton and an anti-soliton can be obtained by evaluating the total energy $E_\parallel+E_\perp$ for the 
$\textrm{K}\overline{\textrm{K}}$ solution (\ref{eq.2sol}) at given times $t_0$ corresponding to a distance $y = 2 v t_0$ between the domain walls. 

Note that $\theta_{2,\textrm{K}\overline{\textrm{K}}}$ is obtained for $g_\perp = 0$, neglecting the feedback of the interchain coupling on the soliton dynamics of the 
spin chain. This approximation is justified in the limit $J_\perp\ll J$ or slightly below the ordering temperature where $M\ll 1$. For larger $g_\perp$ one would have to 
self-consistently determine the soliton solutions in the presence of the mean field from ordered neighboring chains. In this case the equation of motion is 
a double sine-Gordon equation which is not, in general, integrable but nonetheless can be solved numerically.\cite{campbell1,Gani}  

Using the solution $\theta_{2,\textrm{K}\overline{\textrm{K}}}$ of the isolated chain, the confinement potential $V(y) = V_\parallel(y)+V_\perp(y)$ can be computed 
analytically. As a function of the dimensionless separation $\tilde{y} = \sqrt{\kappa} y$ we obtain 
\begin{eqnarray}
\label{eq.Vparallel}
\frac{V_\parallel(\tilde{y})}{E_0} & = & \sqrt{1-\tilde{v}^2} \frac{A(\tilde{y})^{2}}{1+A(\tilde{y})^{2}}\left ( 1+\frac{\mathrm{arcsinh}A(\tilde{y})}{A(\tilde{y})\sqrt{1+A(\tilde{y})^{2}}} \right )\nonumber\\
& & + \frac{1}{\sqrt{1-\tilde{v}^2}}\left (1-\frac{\mathrm{arcsinh}A(\tilde{y})}{A(\tilde{y})\sqrt{1+A(\tilde{y})^{2}}} \right ), \\
\label{eq.Vperp}
\frac{V_\perp(\tilde{y})}{E_0} & = & \frac{g_{\perp}}{\kappa}\sqrt{1-\tilde{v}^2}\frac{A(\tilde{y})\mathrm{arcsinh} A(\tilde{y})}{\sqrt{1+A(\tilde{y})^{2}}}, 
\end{eqnarray}
where we have normalized by the rest energy 
\begin{equation}
\label{eq.E0}
E_0=mc^{2}=2 \rho_{S}\sqrt{\kappa}
\end{equation}
 of a single soliton and defined the function $A(\tilde{y})=\tilde{v}^{-1} \mathrm{sinh}(\tilde{y} /2\sqrt{1-\tilde{v}^{2}})$.

\begin{figure}[t!]
    \centering
    \includegraphics[width=0.9\linewidth]{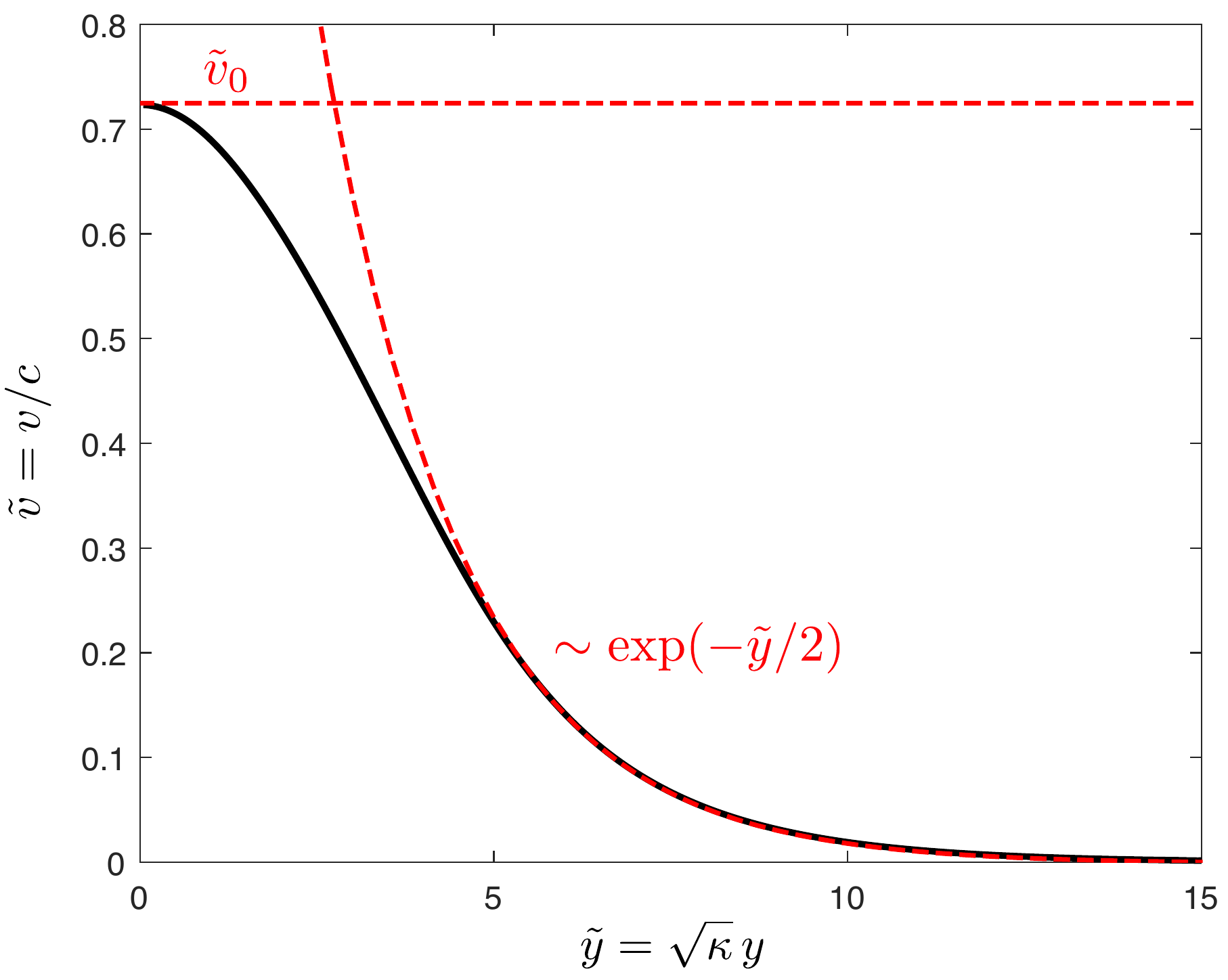}
    \caption{Optimum dimensionless soliton velocity, $\tilde{v}=v/c$, as a function of dimensionless domain wall separation $\tilde{y}$, obtained by minimizing $V_{\parallel}(\tilde{y})$ 
    with respect to $\tilde{v}$.}
    \label{fig2}
\end{figure}

The effective potential still depends on the dimensionless velocity $\tilde{v}$. This parameter can be expressed as a function of the domain-wall separation $\tilde{y}$ 
if we minimize the energy of the isolated chain, $E_\parallel=V_\parallel(\tilde{y})$, with respect to $\tilde{v}$. The resulting function $\tilde{v}(\tilde{y})$ is determined numerically and
plotted in Fig.~\ref{fig2}. While for $\tilde{y}\to 0$ the velocity approaches a constant $\tilde{v}_0\approx 0.725$, for large domain-wall separations the velocity decays exponentially, 
$\tilde{v}\simeq 2.85 \exp(-|\tilde{y}|/2)$.

Let us first investigate the asymptotic behavior of the contributions $V_\parallel$ (\ref{eq.Vparallel}) and $V_\perp$ (\ref{eq.Vperp}) to the potential. At large 
distances ($|\tilde{y}|\to\infty$), the intra-chain contribution $V_\parallel(\tilde{y})$ approaches the energy $2 E_0$ of two free solitons at rest, while the inter-chain 
contribution grows linearly, 
\begin{equation}
\frac{V_\perp (\tilde{y})}{E_0}\approx \frac{g_\perp}{\kappa} |\tilde{y}|.
\end{equation} 

This is the same behavior as for point-like domain walls. This is expected 
since at large distances the spatial extent $\xi$ of the solitons becomes irrelevant. Expressed in terms of the microscopic parameters, using Eqs. (\ref{eq.parameters}), 
(\ref{eq.gperp}) and the definition of $E_0$ (\ref{eq.E0}), we can express the asymptotic result in terms of the microscopic parameters to recover 
$V\sim n_\perp J_\perp |y|/a$. 

At small separations ($\tilde{y}\ll 1$), both contributions are quadratic,
\begin{eqnarray}
\label{eq.Vquadratic}
\frac{V_\parallel(\tilde{y})}{E_0} & \approx & \frac{4-3\tilde{v}_0^2}{6\tilde{v}_0^2\sqrt{1-\tilde{v}_0^2}^3} \tilde{y}^2\approx 2.35 \, \tilde{y}^2,\\
\frac{V_\perp(\tilde{y})}{E_0} & \approx &   \frac{1}{4\tilde{v}_0^2\sqrt{1-\tilde{v}_0^2}} \frac{g_\perp}{\kappa}\tilde{y}^2 \approx 0.69 \frac{g_\perp}{\kappa}\tilde{y}^2,
\end{eqnarray}
which is the result of the gradual annihilation of the extended soliton and anti-soliton. 

\begin{figure}[t!]
    \centering
    \includegraphics[width=0.95\linewidth]{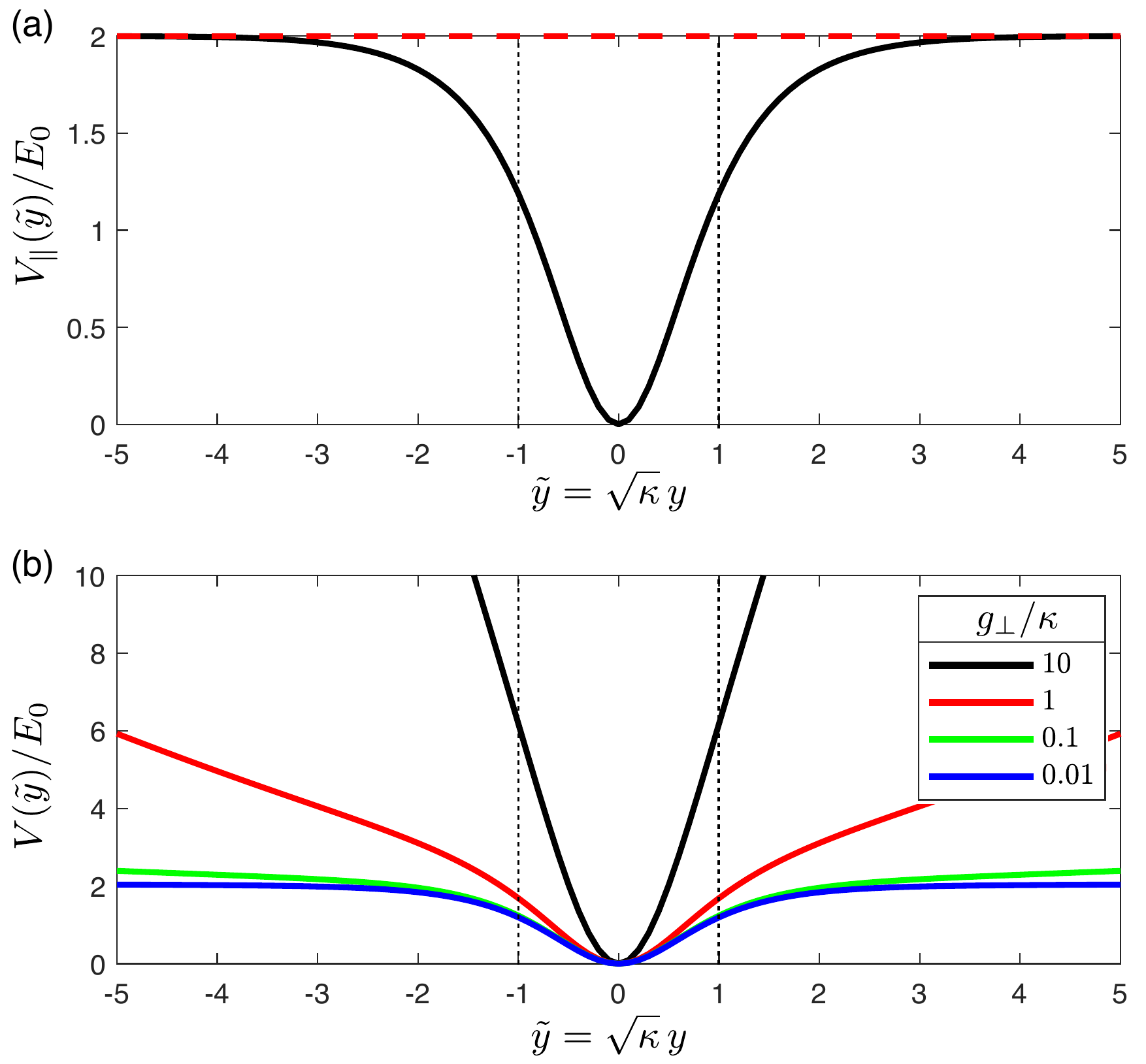}
    \caption{(a) In-chain $\textup{K}\overline{\textup{K}}$ potential $V_\parallel(\tilde{y})$  as a function of dimensionless separation 
    $\tilde{y}=\sqrt{\kappa}y$. At large separations, $V_\parallel$ approaches the energy $2 E_0$ of two free solitons. Due to the gradual destructive interference of the solitons, 
    $V_\parallel$ is rendered quadratic at small distances. The crossover occurs at $\tilde{y}=1$, corresponding to a soliton size $\xi = 1/\sqrt{\kappa}$.
    (b) The same crossover is found in the effective confinement potential $V(\tilde{y})=V_\parallel(\tilde{y})+V_\perp(\tilde{y})$. At large separations the 
    potential is linear, $V(\tilde{y})/E_0\approx (g_\perp/\kappa) |\tilde{y}|$, while at small separations the potential is quadratic due to the gradual annihilation of the extended solitons.}
    \label{fig3}
\end{figure}

The intra-chain contribution $V_\parallel(\tilde{y})$ and the full confinement potential $V(\tilde{y})=V_\parallel(\tilde{y})+V_\perp(\tilde{y})$ are shown in Fig.~\ref{fig3} 
as a function of the dimensionless domain-wall separation $\tilde{y} = \sqrt{\kappa} y$. They display the asymptotic behavior discussed above. The crossover from linear 
to quadratic behavior of $V(\tilde{y})$ occurs at $\tilde{y} =1$. Since the crossover is expected to occur when the solitons start to overlap (see Fig.~\ref{fig1}b), we can 
identify the size of the solitons as 
\begin{equation}
\xi \simeq \frac{1}{\sqrt{\kappa}} = a\sqrt{\frac{J}{2\alpha}}.
\end{equation}

This equation shows that the size of the solitons is controlled by the relative strength of the Ising anisotropy, $\alpha/J$. In the case of strong Ising anisotropy, 
the size of the solitons is of the order of the lattice spacing $a$. On the other hand, in systems with very weak anisotropy, the spatial extent of soliton domain walls can be 
of the order of hundreds of lattice spacings. 
 
\subsection{Bound-State Spectrum}

The gradual destructive interference of extended solitons at separations $y<\xi$ weakens the confinement potential and renders it quadratic.  
In the following we will consider the solitons as point-like particles interacting with the effective non-linear potential $V(y)$ and determine the discrete 
bound-state spectrum from the solution of the one-dimensional Schr\"odinger equation
 \begin{equation}
 \label{eq.Schroedinger}
    -\frac{\hbar^{2}}{2 \mu} \frac{\mathrm{d}^{2}\psi}{\mathrm{d}y^{2}}+V(y)\psi=\epsilon\psi
\end{equation}
for the effective one-body problem for the relative coordinate $y$ of the soliton pair. Here $\mu = m/2$ denotes the reduced mass in terms of the single-soliton mass $m$. 

As a point of reference, let us first consider the limit of very strong Ising anisotropy. In this case the potential is linear down to lattice scale, $V(y)=\lambda |y|$, and the theory 
of linear confinement\cite{Grenier+15,Wang+15,Bera+17,Gannon+19,Coldea+10} applies. The resulting bound-state energies are given by\cite{Wang+15}
\begin{equation}
\label{eq.Airy}
\epsilon_j^> = 2E_0 + \xi_j \lambda^{2/3} \left(\frac{\hbar^2}{\mu}\right)^{1/3},
\end{equation}
where $\xi_j$ are the negative zeroes of the Airy function, $\textrm{Ai}(-\xi_j)=0$, $\xi_1\approx 2.338$, $\xi_2\approx 4.088$, $\xi_3\approx 5.520$, $\ldots$. 

In the limit of very weak anisotropy on the other hand, the confinement potential is quadratic over a significant range, $V(y) \simeq \frac12 \mu\omega^2 y^2$, giving rise
to equidistant energy levels
\begin{equation}
\label{eq.Osc}
\epsilon_j^< = \hbar\omega \left(j+\frac12 \right).
\end{equation}

Due to the crossover of $V(y)$ from quadratic behavior at short distances to linear behavior at large distances, we expect to a related crossover 
in the energy level spacing of the bound states. The strongly bound states at low energies will be almost equidistant, as described by $\epsilon_j^<$ (\ref{eq.Osc}), while the weakly bound 
states at higher energies will approach the sequence $\epsilon_j^>$ (\ref{eq.Airy}). This crossover is controlled by the strength of the Ising anisotropy $\alpha/J$. 

 \begin{figure}[t!]
    \centering
    \includegraphics[width=0.95\linewidth]{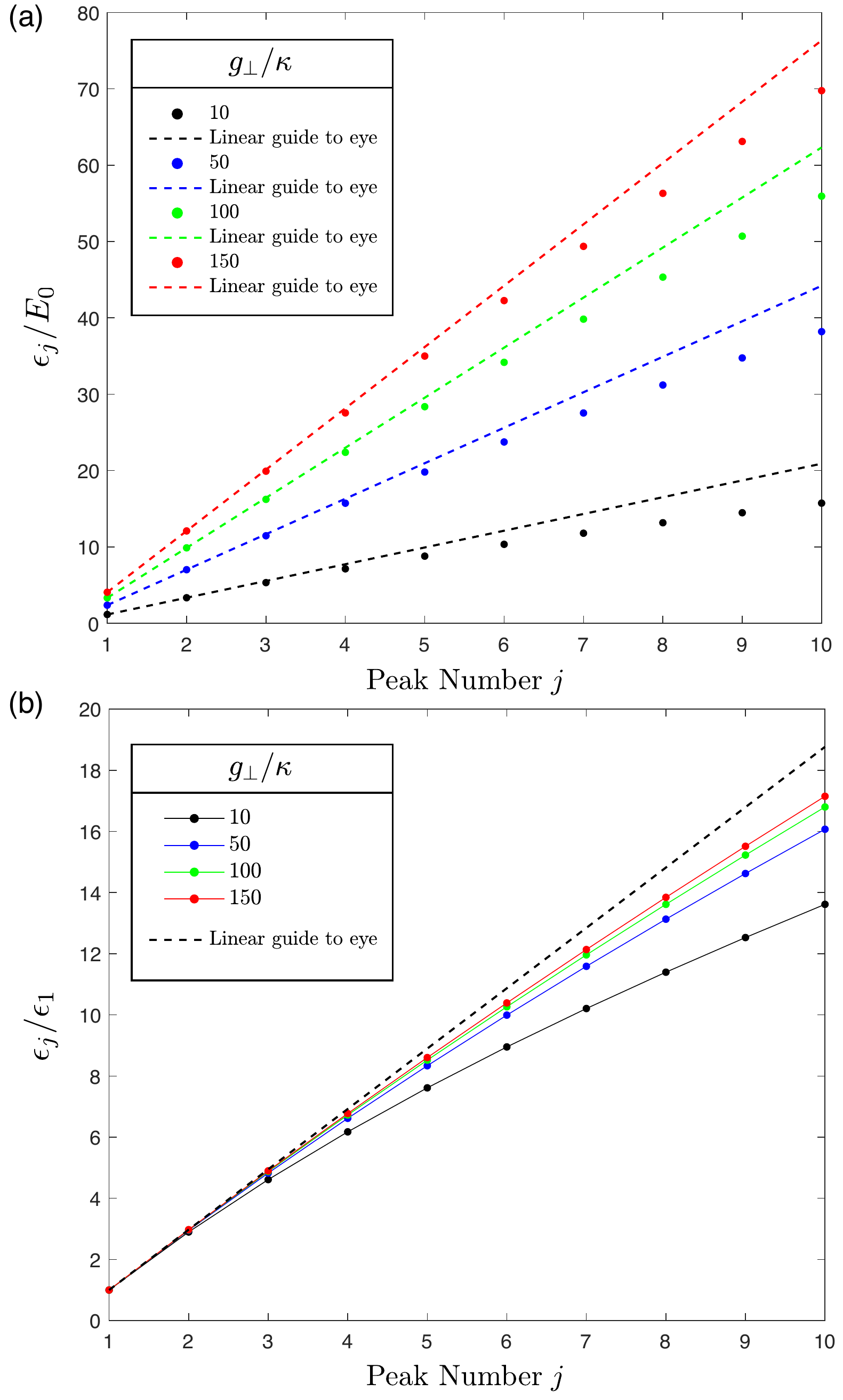}
    \caption{(a) Soliton-antisoliton bound-state energies 
    $\epsilon_j$ in units of the single soliton energy $E_0$  for different ratios $g_\perp/\kappa$ and $S=5/2$. 
    The corresponding confinement potentials are shown in the inset. (b) Same spectra but normalized by the energy $\epsilon_1$ of the first bound state. For larger values of 
    $g_\perp/\kappa$ the level spacing becomes more harmonic oscillator like (dashed line).}
    \label{fig4}
\end{figure}

In order to obtain the bound-state spectrum for the full confinement potential we transform the Schr\"odinger equation (\ref{eq.Schroedinger}) to dimensionless 
units, 
\begin{equation}
 -\frac{1}{2 S^2} \frac{\mathrm{d}^{2}\psi}{\mathrm{d}\tilde{y}^{2}}+\tilde{V}(\tilde{y})\psi=\tilde{\epsilon}\psi,
\end{equation}
$\tilde{y}=\sqrt{\kappa} y$, $\tilde{\epsilon}=\epsilon/E_0$ and $\tilde{V}(\tilde{y})=V_\parallel(\tilde{y})/E_0+V_\perp(\tilde{y})/E_0$ (\ref{eq.Vparallel},\ref{eq.Vperp}), 
and then numerically solve the equation, using the finite-element method implemented in Mathematica.\cite{Mathematica} 

In Fig.~\ref{fig4}(a) the resulting bound-state energies $\epsilon_j/E_0$ for $S=5/2$ (value for CaFe$_{2}$O$_{4}$) and different values of $g_\perp/\kappa$ are shown. 
In the regime of large $g_\perp/\kappa$,  the dominant contribution to the confinement potential comes from the frustrated inter-chain coupling. The tightly 
bound states have almost equidistant energy levels with spacing $\Delta \epsilon/E_0 \approx (1.17/S) \sqrt{g_\perp /\kappa}$, as expected for the asymptotic quadratic form 
of the potential at small distances,  $\tilde{V}(\tilde{y})\approx \tilde{V}_\perp(\tilde{y})\approx 0.69 (g_\perp /\kappa)\tilde{y}^2$. At higher energies, the level spacing 
is reduced because of the crossover of the potential to a linear form at large distances. Normalizing the energies by the energy $\epsilon_1$ of the first bound state 
(see Fig.~\ref{fig4}(b)), it is apparent that the spectrum becomes more like that of a harmonic oscillator if the value of $g_\perp/\kappa$ is increased.


\section{Application to Calcium Ferrite}
\label{sec.experiments}

In this section we will apply our theory of non-linear soliton confinement to the $S=5/2$ antiferromagnet CaFe$_{2}$O$_{4}$. Recent neutron scattering 
experiments\cite{Stock+16} found signatures of solitary magnons in this material with a sequence of nine quantized excitations below the magnetic ordering 
transition at $T_\textrm{N}\approx 200$~K.  

\begin{figure}[t!]
    \centering
    \includegraphics[width=\linewidth]{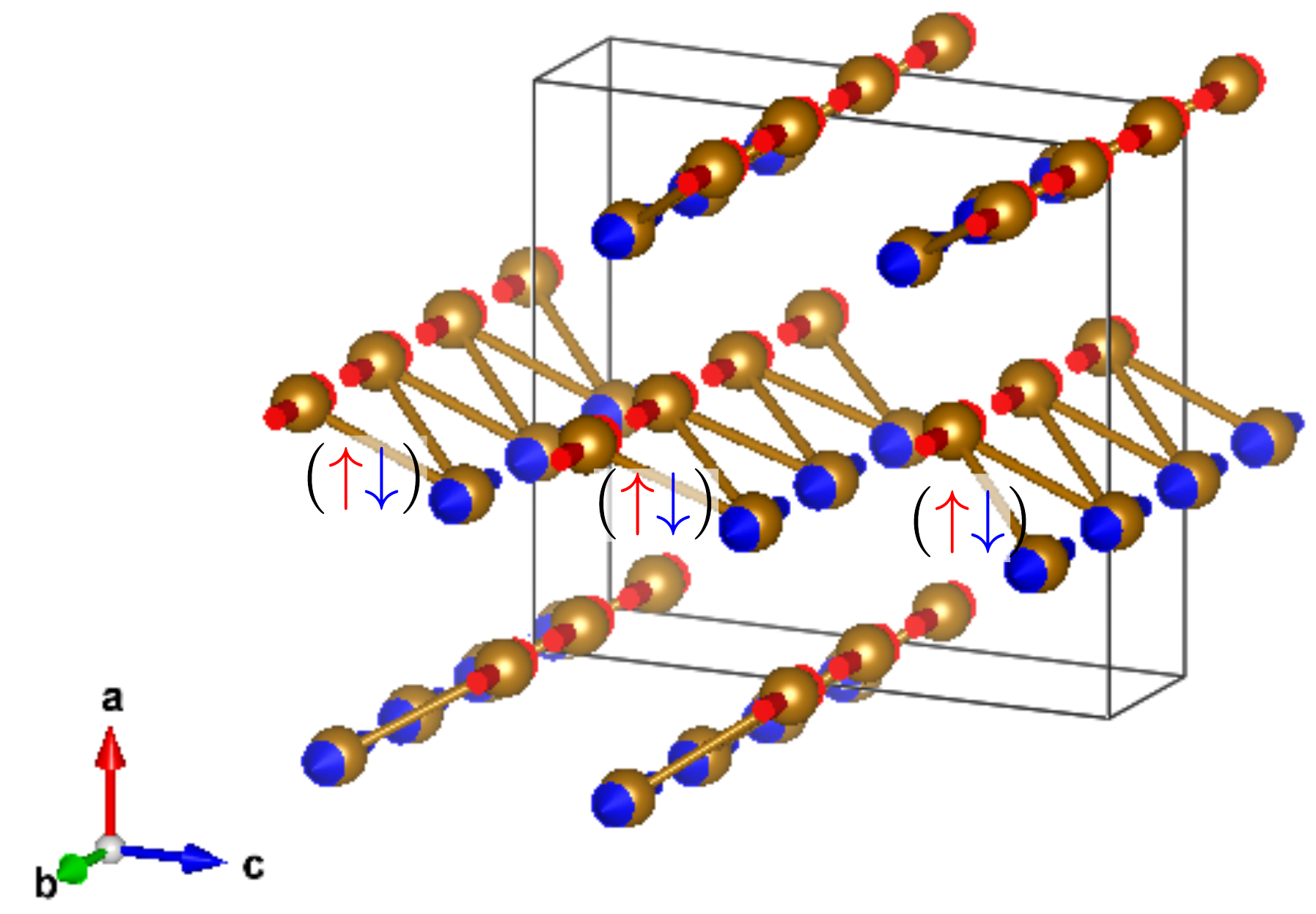}
    \caption{Magnetic structure in the high-temperature B phase of CaFe$_{2}$O$_{4}$,\cite{Stock+16} showing antiferromagnetic zig-zag chains along the $b$ axis. 
    The system exhibits a weak easy-axis anisotropy along $b$. Calcium Ferrite is based upon an orthorhombic unit cell (space group 62 $Pnma$) with dimensions 
    $a=9.230$~\AA, $b=3.017$~\AA, and $c=10.689$~\AA.\cite{Hill+56,Decker+57}}
    \label{fig5}
\end{figure}

CaFe$_{2}$O$_{4}$ has a complex magnetic phase diagram due a competition between two different spin arrangements, termed the $A$ and 
$B$ phases.\cite{Corliss+67} The magnetic structure of the $B$ phase, which dominates at high temperatures,  consists of antiferromagnetic zig-zag chains 
along the $b$ axis (see Fig.~\ref{fig5}). The moments are oriented along $b$ due to a small easy-axis anisotropy. 

The $A$ phase might coexist with the $B$ phase over the full temperature range but becomes clearly visible only below 170~K, which has been identified as its
onset temperature in early studies.\cite{Corliss+67} The two phases are distinguished by their $c$-axis stacking of ferromagnetic $b$-axis stripes: the $B$ phase
consists of stripes with antiferromagnetic alignment within the zig-zag chain, $(\uparrow\downarrow)(\uparrow\downarrow)$, while in the $A$ phase the zig-zag chains
are ferromagnetic with stacking $(\uparrow\uparrow)(\downarrow\downarrow)$ along $c$.\cite{Corliss+67} It has been suggested\cite{Stock+16} that the gradual 
increase of the $A$ phase component is linked to anti-phase domain boundaries along $c$, combined with a continuous change of the Fe-O-Fe bond angle which 
controls the strength and sign of the super-exchange\cite{Mizuno+98,Shimizu+03} between the two legs forming the zig-zag chain. This scenario is supported by the 
presence of diffuse scattering rods along the $L$ direction and spin-wave excitations that show magnetic order  in the $ab$-plane with short-ranged correlations 
along $c$.\cite{Stock+16}

From now on we focus on the $B$ phase that completely dominates at high temperatures where the discrete excitations are observed. As pointed out in Ref.~[\onlinecite{Stock+16}],
the level spacing of the excitations cannot be explained based on the linear confinement picture. This led the authors to speculate 
that the discrete nature of the excitations is not due to interaction-driven bound-state formation but instead a result of spatial confinement  along 
the $c$ axis. Here we show that an effective non-linear interaction potential arising from the extended nature of solitons in CaFe$_{2}$O$_{4}$ would lead to a bound-state spectrum 
that is consistent with the data. 

Let us first inspect the discrete energy-level spectrum presented in Ref.~[\onlinecite{Stock+16}] more closely. The excitations can only be observed above the spin wave anisotropy gap, 
which shows a strong temperature dependence. The gap opens below $T_\textrm{N}\approx 200$~K and  saturates to a value of 
$\Delta\approx 3$~meV below 100~K. For this reason, the lowest energy excitation can only be resolved slightly below $T_\textrm{N}$ where strong fluctuations almost completely 
fill in the gap. The data at 200~K show six discrete energy levels below 2~meV. At 150~K the spin-wave gap almost completely masks this energy range. Instead three 
energy levels become visible above around 1.8~meV. In Ref.~[\onlinecite{Stock+16}] it was assumed that the discrete excitations energies have a negligible temperature dependence and 
that the three levels observed at 150~K are the continuation of the energy sequence at 200~K. 

If the discrete excitations were due to soliton bound-state formation one would expect the excitation spectrum to depend upon temperature. Based on our theory, we expect 
that the main temperature dependence enters through the effective mean-field coupling $g_\perp$ to neighboring chains. Since $g_\perp$ is proportional to the magnetization 
of the system, it increases as temperature is lowered. This would explain why the bound states at 150~K have a  larger level spacing than those at 200~K. 

Moreover, magnetoelastic effects and small changes to the Fe-O-Fe bond angle close to the threshold at which the superexchange would change sign could give rise to 
a non-negligible temperature dependence of magnetic exchange couplings.\cite{Songvilay+20} Finally, the gradual onset of the $A$ phase could 
give rise to additional effects which might obscure the soliton signal in the neutron scattering experiment

\subsection{Experimental Results}

Here we present previously unpublished data that were collected alongside those published in Ref.~[\onlinecite{Stock+16}]. Instead of combining measurements at different temperatures
we focus on $T=200$~K, allowing us to trace the excitations down to very low energies.

Our experiments were performed on single crystals of CaFe$_{2}$O$_{4}$ grown using a mirror furnace. High momentum and energy resolution data was obtained using 
the OSIRIS backscattering spectrometer located at the ISIS Neutron and Muon Source.\cite{Andersen+02}  A white beam of neutrons is incident on the sample and 
the final energy of the scattered neutrons is fixed at $E_f=1.84$~meV using cooled graphite analyzers. A cooled Beryllium filter was used on the scattered side to reduce background. 
The default configuration is  set for a symmetric dynamic range of $\pm 0.5$~meV, however by shifting the incoming energy band width using a chopper the dynamic range was 
extended into the inelastic region. For this experimental setup, the elastic energy resolution (full-width) was $2\delta E=0.025$~meV.
Due to kinematic constraints, we focussed our measurements around 
$\mathbf{Q} =(2, 0, 0)$~(r.l.u) so that the quantized excitations could be tracked up to energy transfers of $\sim 3$~meV.

\begin{figure}[t!]
    \centering
    \includegraphics[width=\linewidth]{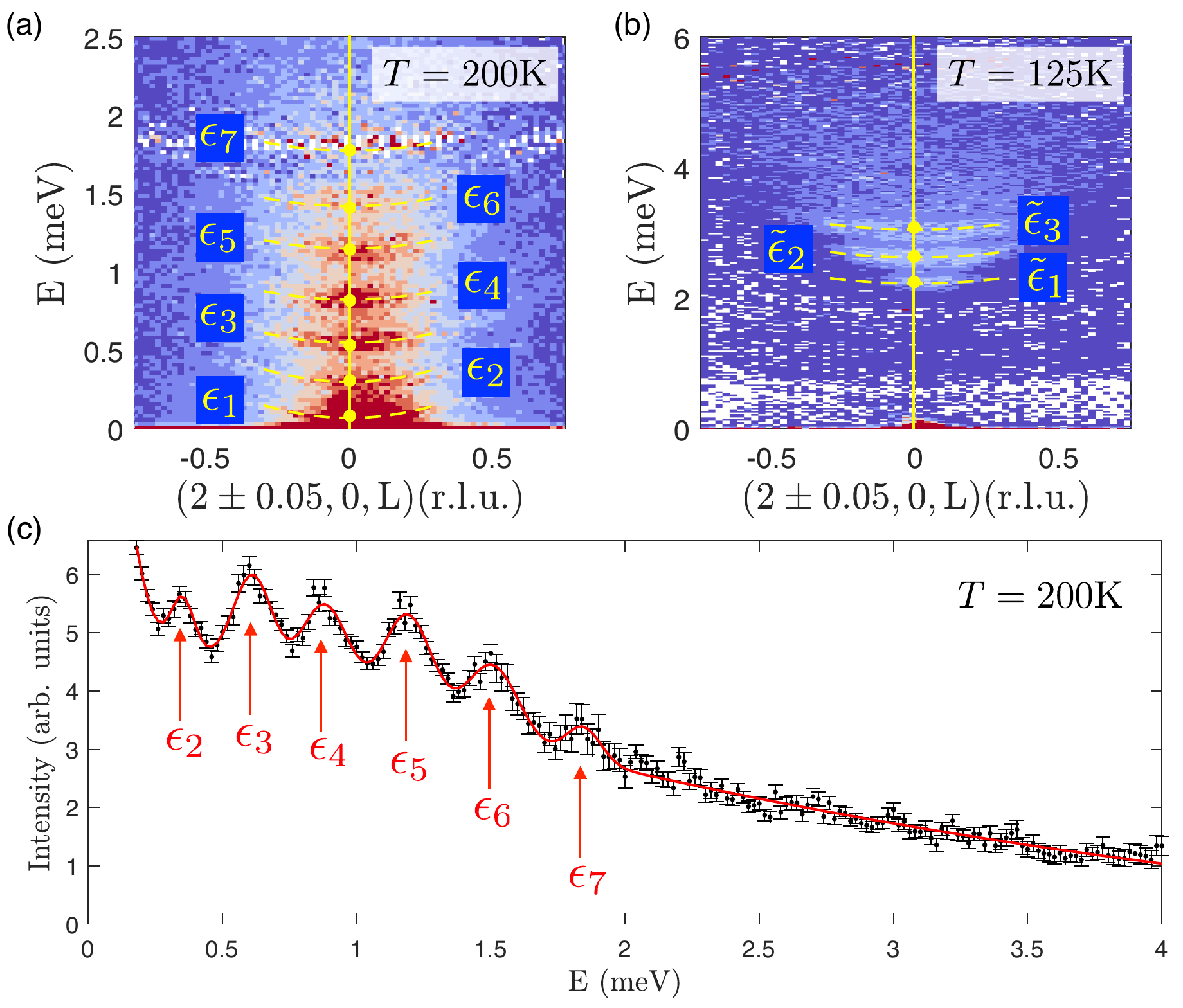}
    \caption{(a) High resolution low energy data recorded on OSIRIS at $T=200$~K, showing seven clearly discernible excitations $\epsilon_j$ at $\mathbf{Q}=(2,0,0)$ (r.l.u.). The modes 
    show a weak quadratic dispersion along $L$, highlighted by dashed yellow lines. (b) At  $T=125$~K the spin-wave gap masks excitations below 2~meV. Above this energy, three 
     additional excitations $\tilde{\epsilon}_j$ are visible. (c) Scattering intensity 
    at $\mathbf{Q}=(2,0,0)$ as a function of energy. Peaks at $\epsilon_2,\ldots,\epsilon_7$ are clearly resolved. The energy $\epsilon_1$ is below the elastic line.}
    \label{fig6}
\end{figure}

As shown in Fig.~\ref{fig6}(a), at 200~K we find seven discrete excitations in low energy scattering data below 2~meV, located at $\mathbf{Q}=(2,0,0)$  and with a weak quadratic 
dispersion along $L$. The intensities are integrated over a small window of $2\pm0.05$ r.l.u. in the $H$ direction. The excitations have an almost linear level 
spacing $\Delta \epsilon\approx 0.3$~meV, in very good agreement with previous results.\cite{Stock+16} 

In comparison, at 125~K the spin-wave gap masks the excitations below 2~meV
but three discrete excitations at $\tilde{\epsilon}_1$, $\tilde{\epsilon}_2$ and $\tilde{\epsilon}_3$ are visible above this energy (Fig.~\ref{fig6}(b)). The modes are at slightly higher 
energies than those identified at 150~K in Ref.~[\onlinecite{Stock+16}], suggesting that there might exists a non negligible temperature dependence.  In the following we will discard the 
excitations above 2~meV since they cannot be resolved at 200~K. 

In Fig.~\ref{fig6}(c), the scattering intensity at 200~K as a function of energy at $\mathbf{Q}=(2,0,0)$  is shown. Peaks at the energy levels $\epsilon_2,\ldots \epsilon_7$ are very clearly 
visible. The first excitation $\epsilon_1$ is beneath the incoherent background in the OSIRIS data and cannot be resolved in the energy cut. However, the energy $\epsilon_1$ can be estimated 
thanks to the weak quadratic dispersion along $L$ (see dashed yellow lines in Fig.~\ref{fig6}(a)).

\subsection{Fitting to Non-Linear Confinement Model}

\begin{figure}[t!]
    \centering
    \includegraphics[width=\linewidth]{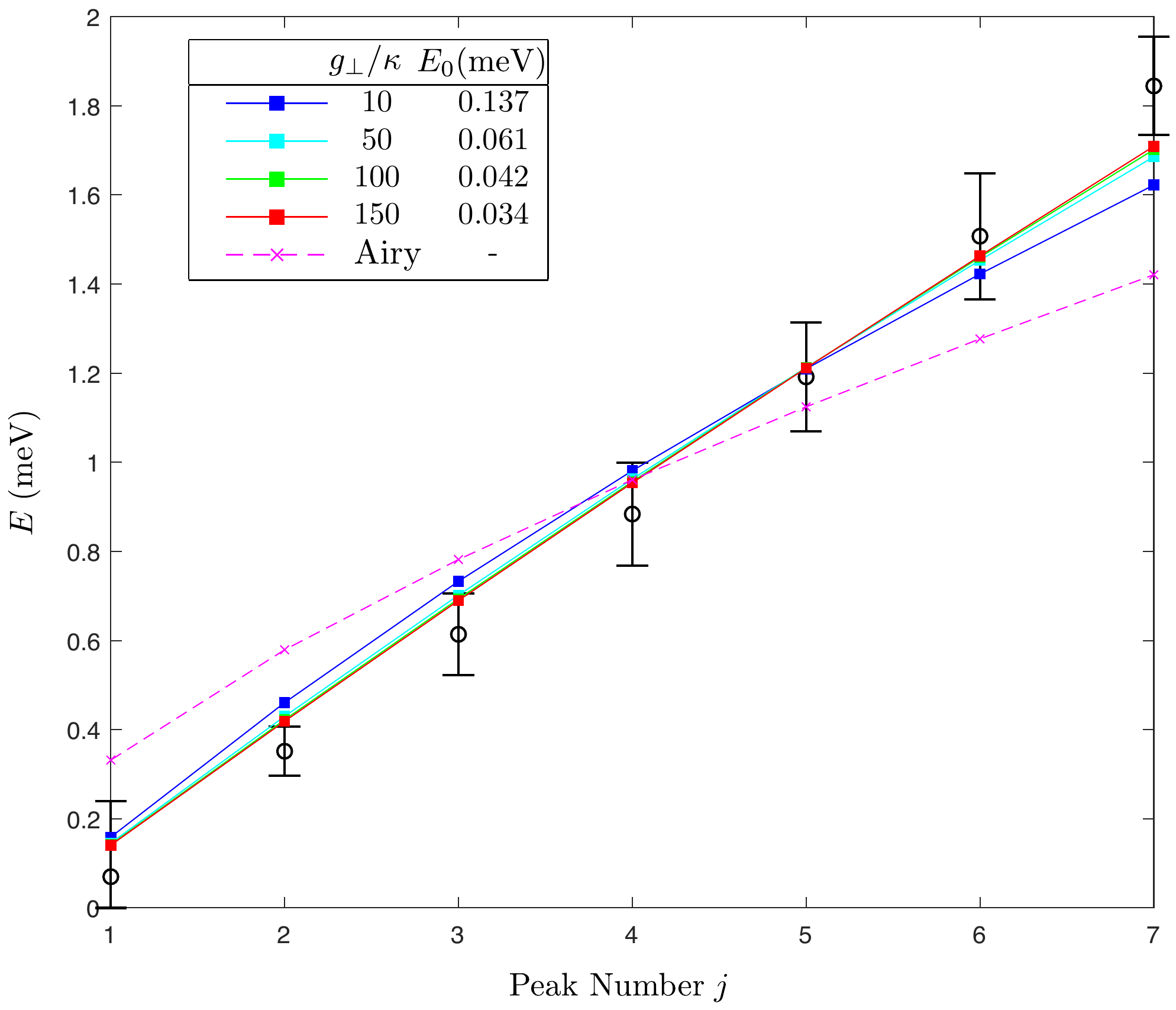}
    \caption{Comparison between the measured excitation energies (open circles) and the soliton bound state energies calculated from the non-linear confinement model. The figure shows 
    best fits to the data for different values of $g_\perp/\kappa$. The quality of the fits improves with increasing values of $g_\perp/\kappa$  and decreasing soliton energy $E_0$. Good agreement 
    is achieved for $g_\perp/\kappa\ge 50$. For comparison, the best fit of the linear-confinement model is shown in magenta.}
    \label{fig7}
\end{figure}

We now investigate whether the seven discrete excitations measured at 200~K can be explained in terms of soliton bound-state formation. The quantized excitations $\epsilon_j$ extracted from 
the neutron scattering experiment are shown as open circles in Fig.~\ref{fig7}. For the levels $j=2,\ldots,7$ we estimate the experimental error $\delta \epsilon_j$
from the full peak width at half maximum. For the lowest energy state, which is masked by the incoherent background of the elastic line, we assume a larger uncertainty of 
$\delta \epsilon_1\approx 0.15$~meV.

 As point of reference, we first assume a linear confinement potential. In this case the soliton bound-state energies would be given by 
$\epsilon_j = A + B\xi_j$, where $\xi_j$ are the negative zeroes of the Airy function and the energies $A$ and $B$ are related to the soliton rest mass and 
the slope of the linear potential, as defined in Eq.~(\ref{eq.Airy}). Here we use $A$ and $B$ as free fitting parameters, not imposing any additional constraints. The resulting best case 
scenario for the linear-confinement model (dashed magenta line in Fig.~\ref{fig7}) strongly deviates from the data, showing that the discrete excitations in CaFe$_{2}$O$_{4}$ 
cannot be understood in terms of a linear confinement of solitons. 

The bound-state spectra obtained from the effective non-linear confinement potential depend on two parameters, the soliton rest energy $E_0=2\rho_S\sqrt{\kappa}$
and the dimensionless ratio $g_\perp/\kappa$. For a given value of $g_\perp/\kappa$ we obtain the best fit to the data $\{\epsilon_j \pm \delta\epsilon_j\}$ by minimizing 
\begin{equation}
\chi^2 = \sum_j \left(\frac{\epsilon_j^\textrm{th}(E_0)-\epsilon_j}{\delta \epsilon_j}\right)^2
\end{equation}
with respect to $E_0$, where $\{\epsilon_j^\textrm{th}\}$  refers to the spectrum obtained from our theoretical model.  

As shown in Fig.~\ref{fig7}, the fits improve with increasing values of $g_\perp/\kappa$, corresponding to decreasing optimum values of $E_0$.  
A good description of our data is obtained for $g_\perp/\kappa=50$ and $E_0=0.061$~meV. Although for larger values of $g_\perp/\kappa$ the fits continue to improve slightly, 
the soliton size $\xi=1/\sqrt{\kappa} = 2\rho_S/E_0$ would  eventually become too large for our theoretical description to be valid.

For $g_\perp/\kappa=50$ the levels are almost equidistant, showing that the first 7 levels fall in the harmonic potential regime. To check consistency, we calculate 
the average mean-square displacement of the soliton bound states, $d_j = \sqrt{ \langle \hat{y}^2 \rangle_j}$, using the approximate quadratic potential (\ref{eq.Vquadratic})
at small distances, $y<\xi$. For the highest level resolved experimentally we obtain $d_7/\xi \approx 0.55 <1$, indicating a significant overlap of the bound solitons. 

The parameters $\rho_S$, $g_\perp$ and $\kappa$ describe the long-wavelength, low energy behavior of the system. This effective continuum description is completely generic and 
applies to any system of weakly coupled antiferromagnetic spin chains in the the large-$S$ limit. 

For illustrative purposes, we have considered a minimal spin model (\ref{eq.Ham}) and established how the effective parameters in the continuum field theory 
depend on the exchange couplings and single-ion anisotropy of the lattice Hamiltonian (see Eqs.~(\ref{eq.parameters}),(\ref{eq.gperp})).  However, this model is too simplistic for 
CaFe$_{2}$O$_{4}$, e.g. it neglects the ferromagnetic exchange along the legs of the zig-zag chains, which is likely to be rather strong. Unfortunately, spin-wave excitations, which could 
be used to determine a more realistic spin model, have not been measured  in the $B$ phase,  but only at 4~K where the competing $A$ phase dominates.\cite{Stock+16} 

On the other hand, close to the N\'eel transition collective fluctuations are very strong, leading to universal behavior detached from microscopic details. 
The spin stiffness is expected to vanish continuously at $T_\textrm{N}$, satisfying Josephson scaling  $\rho_S\sim (T_\textrm{N}-T)^{(d-2)\nu}$ [\onlinecite{Goldenfeld},\onlinecite{Chubukov+94}], 
where $\nu$ is the correlation-length exponent and $d$ the spatial dimension. The bound states are observed slightly below $T_\textrm{N}$ where the stiffness is strongly reduced. 
If we assume $\rho_S/a\approx 3$~meV,  which is of the order of the gap and about a tenth of the spin-wave bandwidth at low temperature, we would obtain a soliton size of about 
100 lattice constants, $\xi/a = 2(\rho_S/a)/E_0\approx 100$.

As suggested in Ref.~[\onlinecite{Stock+16}], quantized excitations in CaFe$_{2}$O$_{4}$ could also arise from anti-phase boundaries along the $c$ axis that separate the two competing 
magnetic phases and lead to spatial confinement. This mechanism is unlikely to be relevant close to $T_\textrm{N}$ where the phase boundaries are dynamic and the $A$ phase is 
almost completely absent. 
At low temperatures, however, the anti-phase domain boundaries become static and carry an uncompensated moment that can be tuned by a 
magnetic field.\cite{Stock+17} The presence of uncompensated spins at phase or domain boundaries is also confirmed by thin-film experiments.\cite{Damerio+20}
Isolated clusters of such orphan spins would provide a natural explanation of the discrete magnetic excitations observed at very low temperatures below the spin-wave 
gap.\cite{Stock+17}


\section{Discussion}
\label{sec.discussion}

To summarize, we have developed a theory for the confinement of solitons in weakly coupled, large-spin antiferromagnetic chains with easy-axis anisotropy. Below the N\'eel transition 
the frustrated interchain coupling generates an attractive potential that leads to the formation of soliton-antisoliton bound states. This mechanism is analogous to the confinement of 
spinons in $S=1/2$  antiferromagnetic XXZ chains\cite{Grenier+15,Wang+15,Bera+17,Gannon+19} or of domain-wall kinks in ferromagnetic Ising chains.\cite{Coldea+10}
But while for these systems the domain-wall defects can be considered as point like, leading to a linear confinement potential, semi-classical solitons have a significant spatial extent. 
This renders the effective confinement potential quadratic on length scales smaller than the size of the solitons, giving rise to a crossover in the energy level spacing of the 
bound states. 

The $S=5/2$ antiferromagnet CaFe$_{2}$O$_{4}$ is a good candidate system to test our theory since this material shows a sequence of discrete low-energy excitations\cite{Stock+16} below 
$T_\textrm{N}$ and exhibits a magnetic structure that  consists of antiferromagnetic zig-zag chains, subject to a weak Ising anisotropy.\cite{Corliss+67} Our inelastic neutron scattering 
experiments, performed slightly below $T_\textrm{N}$,  confirmed the existence of seven discrete excitations below 2~meV with an almost linear level spacing. Our analysis shows
that the quantized excitations can be explained well by the non-linear confinement of large, spatially extended solitons. We argue that strong collective fluctuations close to $T_\textrm{N}$ 
 play a crucial role, collapsing the anisotropy gap and strongly reducing the spin stiffness.  

There are many possible ways in which our theory can be extended to describe a rich variety of physical systems. To model materials with strong interchain coupling one can include 
the feedback of the effective field from neighboring chains on the soliton dynamics. Such a staggered field changes the equation of motion to a double sine-Gordon equation
which is no longer integrable but nonetheless can be solved numerically.\cite{campbell1,Gani} Staggered fields could also be generated by applying external fields in systems with 
staggered $g$ tensors.\cite{Oshikawa+97,Affleck+99} Since solitons and antisolitons have opposite chirality it would be interesting to study the effects 
of a weak Dzyaloshinskii-Moriya interaction which would introduce chirality in the antiferromagnetic background. Finally, one might include finite-lifetime effects due to collisions of  
bound soliton pairs and the interactions with spin-wave excitations.

Thanks to recent advances in crystal growth and neutron scattering technology it is now possible to resolve soliton bound states at very low energies.
 The relevant theoretical parameters in the effective long-wavelength description, such as the spin stiffness, spin-wave velocity and staggered magnetization, 
vanish at the continuous N\'eel transition, showing characteristic power-law behavior. The measurement of soliton bound states close to the transition could therefore 
provide a novel route to study universal critical behavior in inelastic neutron scattering experiments.

{\bf{Acknowledgements}}
The authors thank E. Christou,  A. Green, A. James and M. Songvilay for useful discussions. F.K. acknowledges financial support from EPSRC under Grant No. EP/P013449/1.
S.W.C was supported by the DOE under Grant No. DOE: DE-FG02-07ER46382.
C.S. and H.L. wish to thank the EPSRC and the STFC for funding.
H.L. was co-funded by the ISIS facility development studentship programme. Experiments at the ISIS Neutron and Muon Source were supported by a beamtime allocation 
RB1510445 (DOI: 10.5286/ISIS.E.RB1510445) from the Science and Technology Facilities Council.

\end{document}